\documentclass[onecolumn]{revtex4-2}

\usepackage{enumerate}
\usepackage{enumitem}
\usepackage{graphicx}
\usepackage{dcolumn}
\usepackage{bm}
\usepackage{graphicx}
\usepackage{amssymb}
\usepackage{epstopdf}
\usepackage{color}

\usepackage{soul}

\begin{document}
\title{\Large \textbf{Gate-tunable hallmarks of unconventional superconductivity in non-centrosymmetric nanowires }}

\author{Gyanendra Singh,$^{1,*}$ Claudio Guarcello,$^{2,3}$ Edouard Lesne,$^4$ Dag Winkler,$^1$ Tord Claeson,$^1$ Thilo Bauch,$^1$ Floriana Lombardi,$^1$ Andrea D. Caviglia,$^4$ Roberta Citro,$^{2,3,5}$ Mario Cuoco,$^{5,*}$ Alexei Kalaboukhov$^{1,*}$ }

\affiliation{$^1$ Department of Microtechnology and Nanoscience - MC2, Chalmers University of Technology, SE 412 96 Gothenburg, Sweden. \\
$^2$ Dipartimento di Fisica “E. R. Caianiello,” Università di Salerno, Via Giovanni Paolo II 132, I-84084 Fisciano (SA), Italy. \\
$^3$ INFN, Sezione di Napoli Gruppo Collegato di Salerno, Complesso Universitario di Monte S. Angelo, I-80126 Napoli, Italy.\\
$^4$ Kavli Institute of Nanoscience, Delft University of Technology, Lorentzweg 1, 2628 CJ Delft
The Netherlands. \\
$^5$ SPIN-CNR, c/o Università di Salerno, I-84084 Fisciano (SA), Italy.}
\email[]{singhgy@chalmers.se, mario.cuoco@spin.cnr.it, alexei.kalaboukhov@chalmers.se}

\vskip 5cm

\begin{abstract}
\textbf{\normalsize
Two dimensional SrTiO$_3$-based interfaces stand out among non-centrosymmetric superconductors due to their intricate interplay of gate tunable Rashba spin-orbit coupling and multi-orbital electronic occupations, whose combination theoretically prefigures various forms of non-standard superconductivity. However, a convincing demonstration by phase sensitive measurements has been elusive so far. Here, by employing superconducting transport measurements in nano-devices we present clear-cut experimental evidences of unconventional superconductivity in the LaAlO$_3$/SrTiO$_3$ interface. The central observations are the substantial anomalous enhancement of the critical current by small magnetic fields applied perpendicularly to the plane of electron motion, and the asymmetric response with respect to the magnetic field direction. These features have a unique trend in intensity and sign upon electrostatic gating that, together with their dependence on temperature and nanowire dimensions, cannot be accommodated within a scenario of canonical spin-singlet superconductivity. We theoretically demonstrate that the hallmarks of the experimental observations unambiguously indicate a coexistence of Josephson channels with sign difference and intrinsic phase shift. The character of these findings establishes the occurrence of independent components of unconventional pairing in the superconducting state due to inversion symmetry breaking. The outcomes open new venues for the investigation of multi-orbital non-centrosymmetric superconductivity and Josephson-based devices for quantum technologies.}
\end{abstract}
\maketitle

Accessing the fundamental structure of Cooper pairs in unconventional superconductors~\cite{Sigrist-1991}
and the mechanisms behind electron pairing are among the most notable challenges in condensed matter physics. A superconductor exhibits unconventional pairing when one has breaking of either time reversal or inversion symmetries apart from the canonical gauge symmetry. Two-dimensional (2D) superconductors offer new opportunities to study and tailor unconventional superconducting order parameter as they inherently lack inversion symmetry in the presence of Rashba-type spin orbit coupling and allow large electrostatic tuning of charge carrier density and bands occupation~\cite{Nam-2016, Saito-2016, Cao-2018, Sohn-2018}.
The 2D electron systems in SrTiO$_3$ based heterostructures, such as LaAlO$_3$/SrTiO$_3$ (LAO/STO)~\cite{Ohtomo-2004, Reyren-2007}, are a versatile platform for studying non-centrosymmetric multi-orbital superconductivity~\cite{Fukaya-2018, Scheurer-2015-2, Fernandes-2013} due to the ability to modulate by electrostatic gating the superconducting critical temperature~\cite{Caviglia-2008,Thierschmann-2018,Hurand-2015} together with the strength of the Rashba spin-orbit coupling~\cite{Caviglia-2010,BenShalom-2010} and the occupation of the Ti 3d orbitals (d$_{xy}$,d$_{xz}$,d$_{yz}$)~\cite{Joshua-2012, Herranz-2015}.
Remarkably, the combination of inversion symmetry breaking and multiple orbital degrees of freedom can yield a superconducting order parameter that goes beyond the canonical singlet-triplet mixed parity, with an inter-band anti-phase pairing (e.g., $s_{+-}$)~\cite{Scheurer-2015-2} or pure even-parity inter-orbital spin-triplet pairs~\cite{Fukaya-2018}.
Recently, few experimental evidences have been already considered as a direct manifestation of an $s_{+-}$ pairing or as a generic indication of an unconventional type of superconductivity. Notable examples are the observation of a superconducting gap suppression nearby a Lifshitz transition~\cite{Singh-2019,Trevisan-2018}, the anomalous magnetic field dependence of critical current in weak links~\cite{Bal-2015,Stornaiuolo-2017} and uniform nanowires~\cite{Kalaboukhov-2017}, and several in-gap bound states probed by tunneling spectroscopy~\cite{Kuerten-2017}.

\begin{figure*}[t]
\begin{center}
\includegraphics [width=18cm]{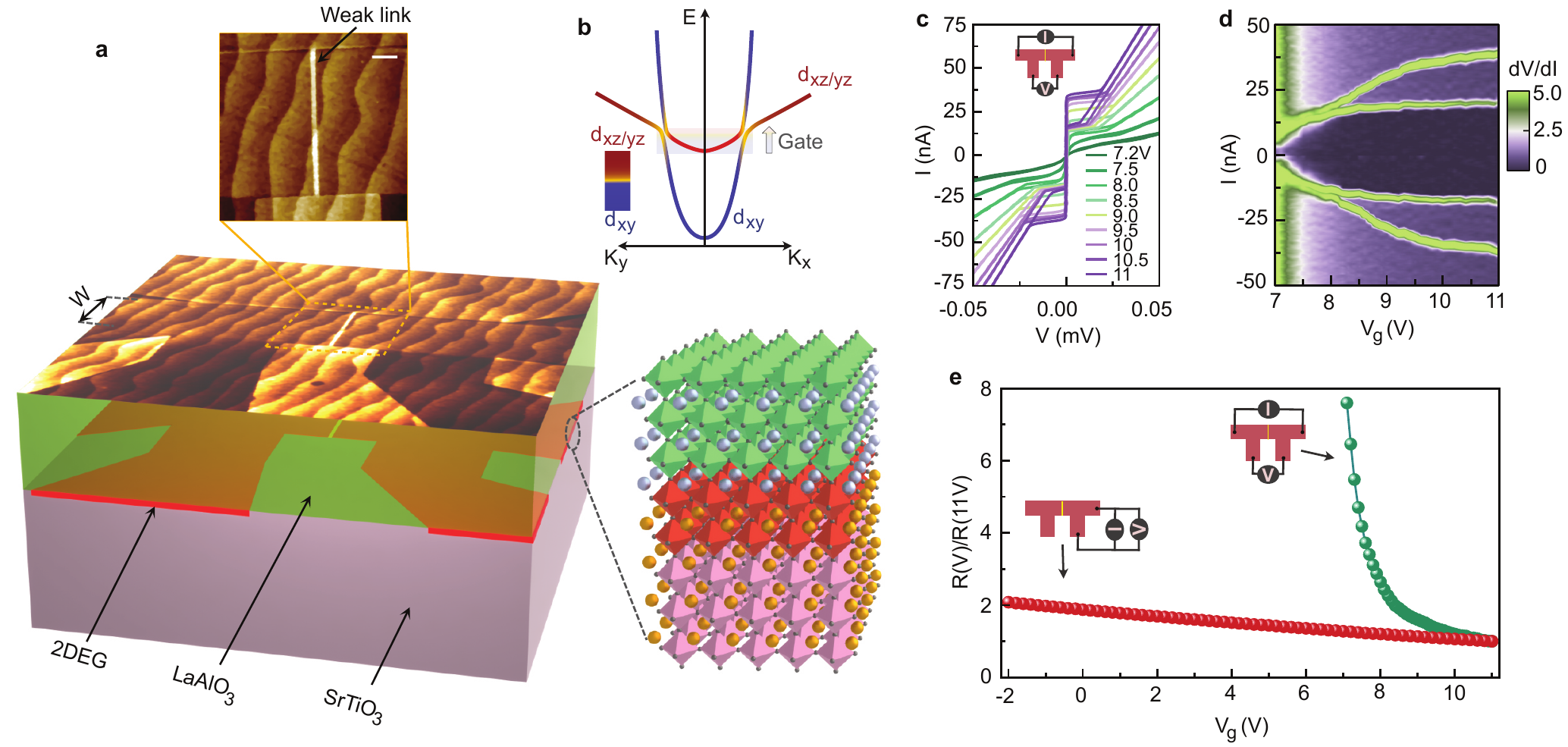}%
\end{center}
\vskip -0.5cm
\caption{\label{fig1} \textbf{Device geometry and characterization.} \textbf{a,} Schematic of LaAlO$_3$/SrTiO$_3$ device fabricated using double step electron beam lithography and low energy Ar+ ion irradiation. The upper surface of the schematic represents the actual atomic force microscopy topographic scan of the device. The dark region represents the conducting interface, while the light area corresponds to the insulating interface. We identified width $W \sim1 \;\mu\text{m}$ for this device. The upper panel shows a blow-up of the weak link region in the middle of the device with weak link width $d \sim35\;\text{nm}$. Scale bar, 200 nm. The right panel shows a schematic of the crystal structure. \textbf{b,} Sketch of band structure of the 2DEG, adapted from Ref. 10. \textbf{c,} Current $I$ \emph{vs} voltage $V$ traces at different gate voltage $V_{g}$ measured at $20\;\text{mK}$ in a four-probe configuration. \textbf{d,} The gate voltage-dependent color plot of $dV/dI$. \textbf{e,} The normalized resistance \emph{vs} gate voltage curve obtained from the fitting of the linear region of I-V’s—the curves measured in two different geometry presented in the inset. The green curve is the measured resistance using a four-probe method where current flows through the weak link while red data are obtained in two probe configuration where the weak link excluded.}
\end{figure*}

Additionally, there are also definite indications of intrinsic inhomogeneities in the normal~\cite{Kalisky-2013, Goble-2017} and superconducting~\cite{Bert-2011, Caprara-2015, Singh-2018} states of the LAO/STO interface originating either from structural domains or electronic phase separation~\cite{Biscaras-2013}.
Therefore, the superconducting phase is likely to be marked by nonuniform superconducting islands~\cite{Singh-2018,Hurand-2019,Nicola-2019}.
While these strong nonmagnetic inhomogeneities at first sight can be incompatible with an unconventional type of pairing, it has been recently figured out that extra internal
degrees of freedom such as orbitals, sublattices, or valleys can protect Cooper pairs from strong scattering~\cite{Andersen-2020}.
Then, local inter-band anti-phase pairing or inter-orbital spin-triplet states are not precluded and can provide robust features of interference effects even in inhomogeneous superconducting conditions. The overall scenario underlines fundamental challenges not yet fully settled about the unconventional nature of the superconducting state in LAO/STO interface, its interrelation with the multi-orbital degrees of freedom, the survival in a strongly inhomogeneous environment, as well as the eventual occurrence of extra symmetry breaking.

In this work, we performed a detailed investigation of the superconducting transport in LAO/STO nano-devices as a function of magnetic field and gate voltage. The main finding is very strong anomalous enhancement of critical current ($I_c$) in small perpendicular magnetic field ($B_\perp$). The effect is significantly enhanced at positive gate voltages corresponding to the regime of multi-band occupation. This anomalous magnetic field dependence of critical current cannot be explained by classical models of $s$-wave superconductivity.
We propose a phenomenological model based on multiband non-centrosymmetric superconductor in which the enhancement of critical current is explained as an interference effect of Josephson channels with intrinsic phase shifts. The symmetry of $I_c$($B_\perp$) response excludes time reversal symmetry breaking mechanism for the occurrence of unconventional pairing components. We show that inversion symmetry breaking and non-trivial multi-orbital superconductivity is sufficient to account for the experimental observations and gate tunability of the effect.

\begin{figure}[t]
\begin{center}
\includegraphics [width=12cm]{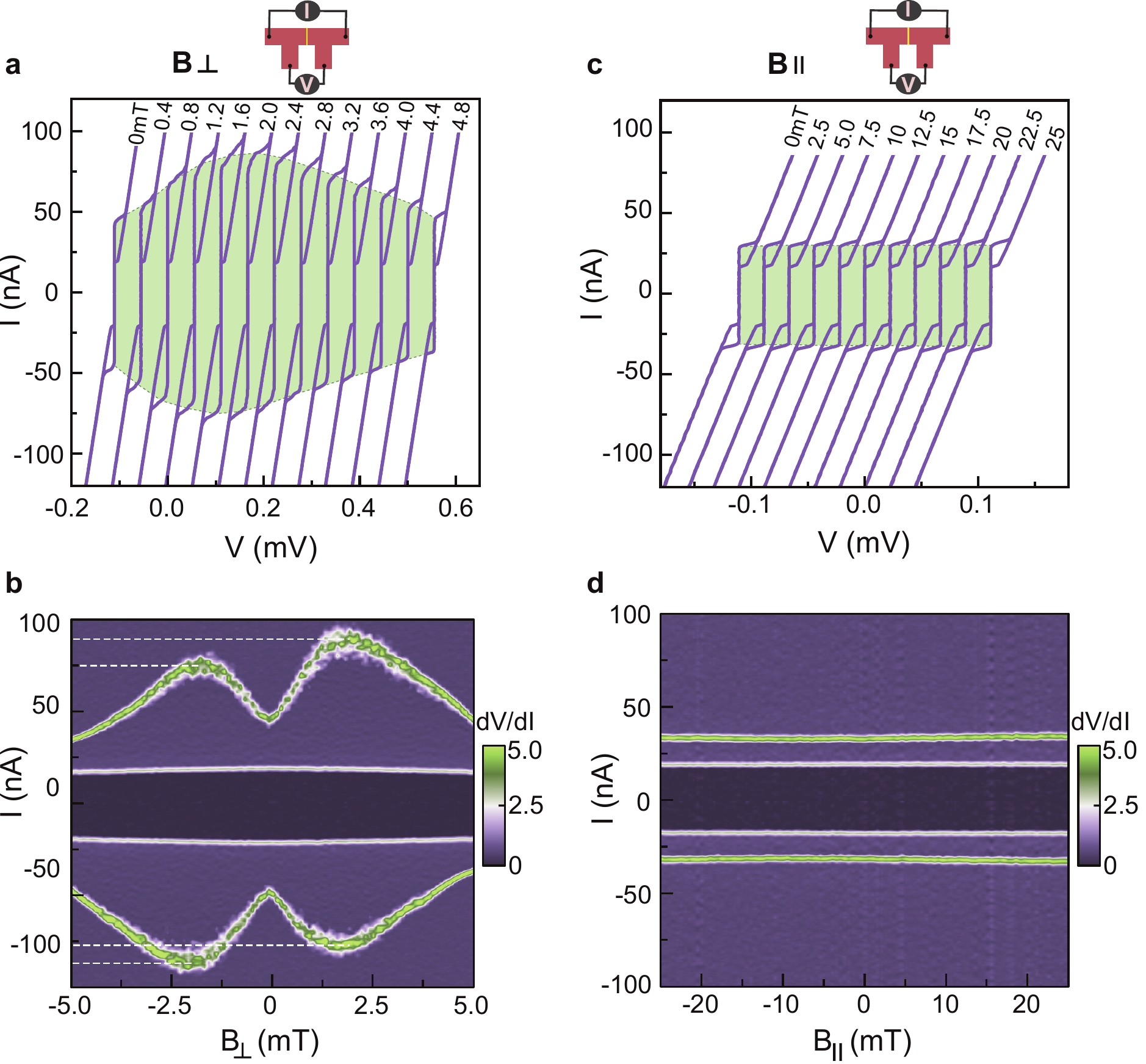}%
\end{center}
\vskip -0.5cm
\caption{\label{fig2} \textbf{Evolution of critical current with magnetic field.} \textbf{a,} Selected current $I$ \emph{vs} voltage $V$ traces of the device measured at $11\;\text{V}$, and $20\;\text{mK}$ as a function of the magnetic field, in the range of $\pm 5\;\text{mT}$, applied perpendicular to the plane of the sample. The curves are shifted horizontally for clarity. A background color highlights critical current variation with $B_\perp$. \textbf{b,} Contour plot of $dV/dI$ as a function of the current and the out-of-plane magnetic field. \textbf{c,} Selective I-V traces measured at $11\;\text{V}$, and $20\;\text{mK}$ as a function of in-plane magnetic field applied the range of $\pm25\;\text{mT}$. \textbf{d,} Contour plot of $dV/dI$ as a function of the current and the in-plane magnetic field.}
\end{figure}

The devices in this study are bridges with width of $1\;\mu$m that contain artificially created weak links with barrier length of $35\;\text{nm}$, see Fig. 1a. Devices were patterned in 5 uc LaAlO$_3$ thin film deposited on TiO$_2$ terminated (001)-SrTiO$_3$ substrate using double-step electron-beam lithography (EBL) and low energy ion irradiation technique, see Methods section. To create the weak links, we utilized the unique property of Ar+ ion irradiaion that allows gradual tuning of interface resistance~\cite{Singh-2021}.

The properties of weak links were first investigated by measurements of their I-V characteristics at $T = 20\;\text{mK}$ and different gate voltages ($V_g$) applied at the back side of the SrTiO$_3$ substrate (Fig. 1b). The I-V characteristics were obtained by measuring voltage drop in a four-terminal configuration across the bridge. There is an apparent hysteresis in the I-V characteristics at high gate voltages characterized by the presence of high critical current, $I_c$, and small return current, $I_{cr}$. This hysteresis in nanowires is often attributed to Joule heating~\cite{Tinkham-2003, Kalaboukhov-2017}. The critical current $I_c$ decreases non-monotonically with reducing back-gate voltage from $11\;\text{V}$ to $7.2\;\text{V}$ (Fig. 1c) due to the loss of superconducting phase coherence. At the same time, the return current is gate voltage-independent that further supports its thermal origin. The weak links are clearly discerned from the uniform bridges by their increased normal resistance (Fig. 1d) and much faster decay of critical current as a function of the gate voltage (Supplementary Fig. S7). This implies that the transparency of our artificial weak link is reduced faster as compared to the one of the electrode.

Next, we present the effect of the applied out-of-plane magnetic field ($B_\perp$) on the critical current. Magnetic field normally suppresses the critical current due to the pair breaking caused by orbital or paramagnetic effects. In contrast, we observe a substantial enhancement of critical current $I_c$ as soon as a small $B_\perp$ is applied (Fig. 2a). The $I_c$ increases from $\sim35\;\text{nA}$ at $0\;\text{mT}$ to $\sim 90\;\text{nA}$ at $+2\;\text{mT}$ and then shows a gradual reduction for further increase of $B_\perp$. Plotting $dV/dI$ as a function of $B_\perp$ and current reveals a striking asymmetric dependence of $I_c$ as a function of direction of $B_\perp$ (Fig. 2b). The asymmetry results in different amplitudes of positive ($I_c^+$) and negative ($I_c^-$) values of critical current at finite magnetic field. However, the magnetic field pattern is symmetric with respect to the change of sign of both critical current and magnetic field, i.e. $I_c^+(B_\perp) = - I_c^- (-B_\perp)$, indicating the absence of time-reversal symmetry breaking.

We also note that the effect is observed only when magnetic field is applied perpendicularly to the plane, and there is no enhancement of $I_c$ in the parallel magnetic field ($B_\parallel$) as shown in Fig. 2c,d. The normal resistance does not show magnetic field dependence similar to $I_c(B_\perp)$, see Supplementary information Fig. S8, indicating that the effect is due to enhancement of superconducting current. There is also negligible hysteresis in $I_c(B_\perp)$ dependence that rules out the presence of intrinsic magnetism in our samples, in contrast with earlier reports where very strong hysteresis in magnetoresistance was observed \cite{Ron-2014}.

An identical enhancement of $I_c$ between $0-2\;\text{mT}$ is also observed in uniform bridges without artificial weak link (Supplementary Fig. S8), but with a relatively large maximum critical current, $I_c (2\;\text{mT}) \sim130\;\text{nA}$, and also a clear asymmetric response as a function of the $B_\perp$ direction. This is in agreement with the presence of intrinsic weak links in the interface~\cite{Caprara-2015, Singh-2018,Biscaras-2013}. The artificial weak link allows much more effective tunability of barrier transparency as a function of gate voltage as compared with electrodes and thus serves as a better probe of the order parameter.
\begin{figure*}[t]
\begin{center}
\includegraphics [width=17cm]{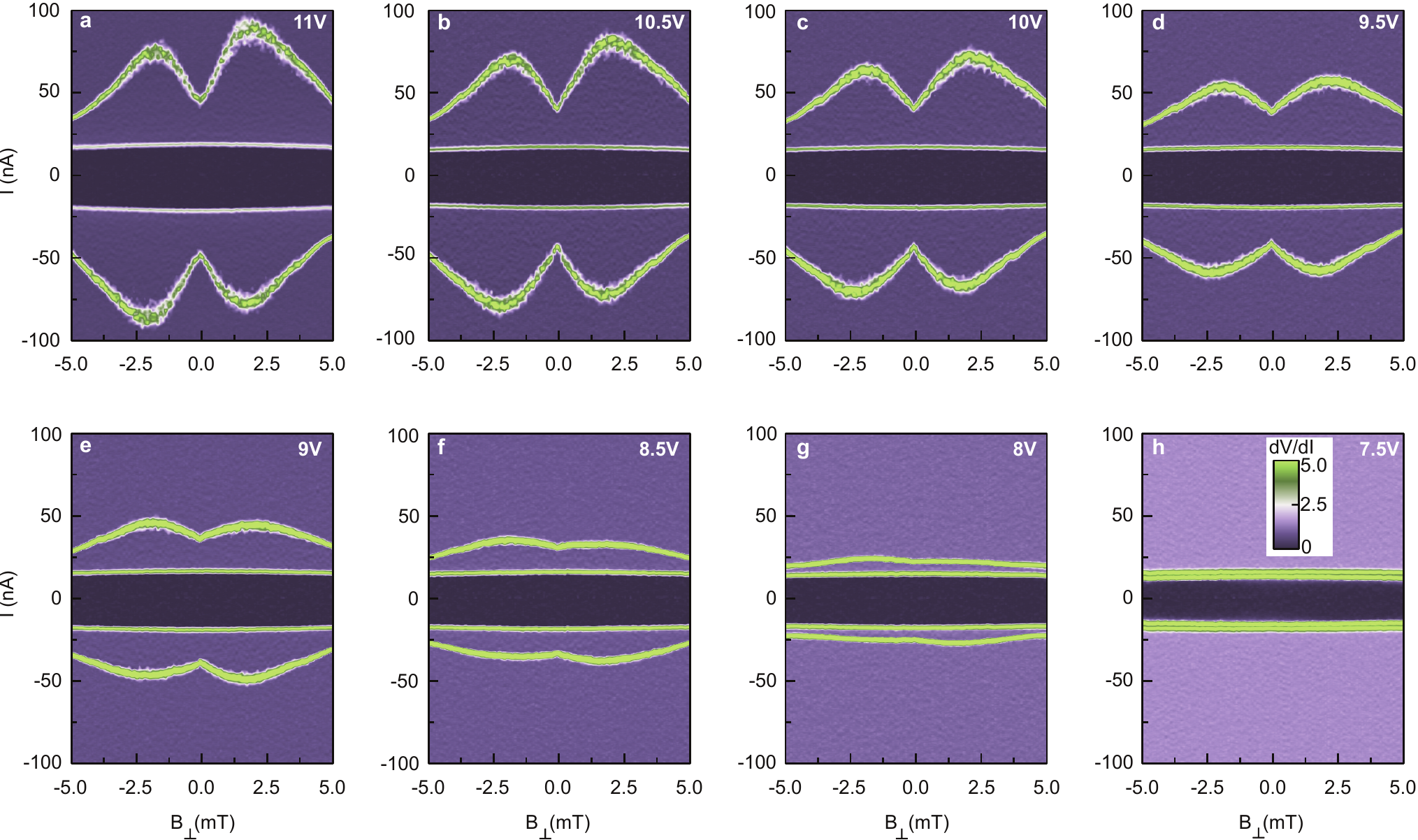}%
\end{center}
\vskip -0.5cm
\caption{\label{fig3} \textbf{Gate voltage effect on critical current enhancement.} \textbf{a-h,} the contour plot of $dV/dI$ as a function of the current and out-of-plane magnetic field measured across the weak link at $20\;\text{mK}$ and constant gate voltages between $11\;\text{V} - 7.5\;\text{V}$.}
\end{figure*}
The effect of $I_c$ enhancement as a function of $B_\perp$ is strongly suppressed as $V_g$ is reduced and completely disappears below 7.5$\;\text{V}$, see Fig.3 a-h. In Fig. 4a, we compare $I_c$ as a function of the gate voltage estimated at the minimum ($B_\perp=0\;\text{mT}$) and maximum ($B_\perp=2\;\text{mT}$) of $I_c$ obtained from the curves in Fig. 3. It is readily apparent that the $I_c(2\;\text{mT})$ shows more rapid suppression with $V_g$ in comparison to smooth reduction at $0\;\text{mT}$, see also Supplementary figure S9 for the evolution of $I_c$ with $V_g$ of uniform bridge. There is also a sign reversal in asymmetry with respect to the magnetic field direction at $\sim9.5\;\text{V}$, as evident from Fig. 4b, where we show the normalized positive critical current as a function of $B_\perp$ at different $V_g$. In Fig. 4c, $\Delta I_n=I_n^+(2\;\text{mT})-I_n^-(-2\;\text{mT})$ is plotted as a function of $V_g$, where $I_n^+(I_n^-)$ represents the value of normalized positive critical current at $2\;\text{mT}$ ($-2\;\text{mT}$), which exhibits a transition from positive to negative values at $9.5\;\text{V}$. We performed similar $I_c(B_\perp)$ measurements in the gate voltage range of $11\;\text{V}$ to $-2\;\text{V}$ for uniform bridges, as shown in Supplementary Fig. S9. The result, in this case, does not exhibit clear sign change in the asymmetry of $I_c(B_\perp)$. Therefore, the behavior of the asymmetry in $I_c(B_\perp)$ is characteristic of the artificial weak link.

The critical question to start with is whether the observed enhancement of $I_c$ can be understood within classical superconductivity frameworks, or one needs to invoke scenarios with unconventional pairing. There are various effects proposed earlier to explain the anomalous enhancement of the critical current by magnetic fields, such as those due to an internal compensation of intrinsic exchange fields by the external magnetic field~\cite{Jaccarino-1962}. This mechanism however is also expected to enhance the critical temperature as a function of the applied magnetic field, as observed in the previous reports~\cite{Gardner-2011}. To evaluate this compensation mechanism, we have measured the I-V’s as a function of $B_\perp$ at different temperatures below the critical temperature, and resistance \emph{vs} $B_\perp$ in the vicinity of the superconducting transition. In Supplementary Fig. S6a-h, we show the I-V($B_\perp$) measured at different temperatures and at $11\;\text{V}$. The amplitude of $I_c$ enhancement reduces with increasing the temperature and disappears above $70\;\text{mK}$, much below the critical temperature of the device. In addition, the critical temperature estimated from $R(T)$ measurements (Supplementary Fig. S6i) displays a linear decrease with the $B_\perp$ (Supplementary Fig. S6j). Thus, this scenario is incompatible with our results, which only display a pronounced effect at the lowest temperature below the superconducting transition ($T_c$ $\sim$ 105mK). We can also rule out the effect of suppression of spin-flip scattering on magnetic impurities~\cite{Kharitonov-2001} as it is expected to be enhanced when the magnetic field is aligned parallel to the interface~\cite{Rogachev-2006, Wei-2006}, which is not consistent with our experimental results, see Fig. 2c,d.
\begin{figure*}[t]
\begin{center}
\includegraphics [width=17cm]{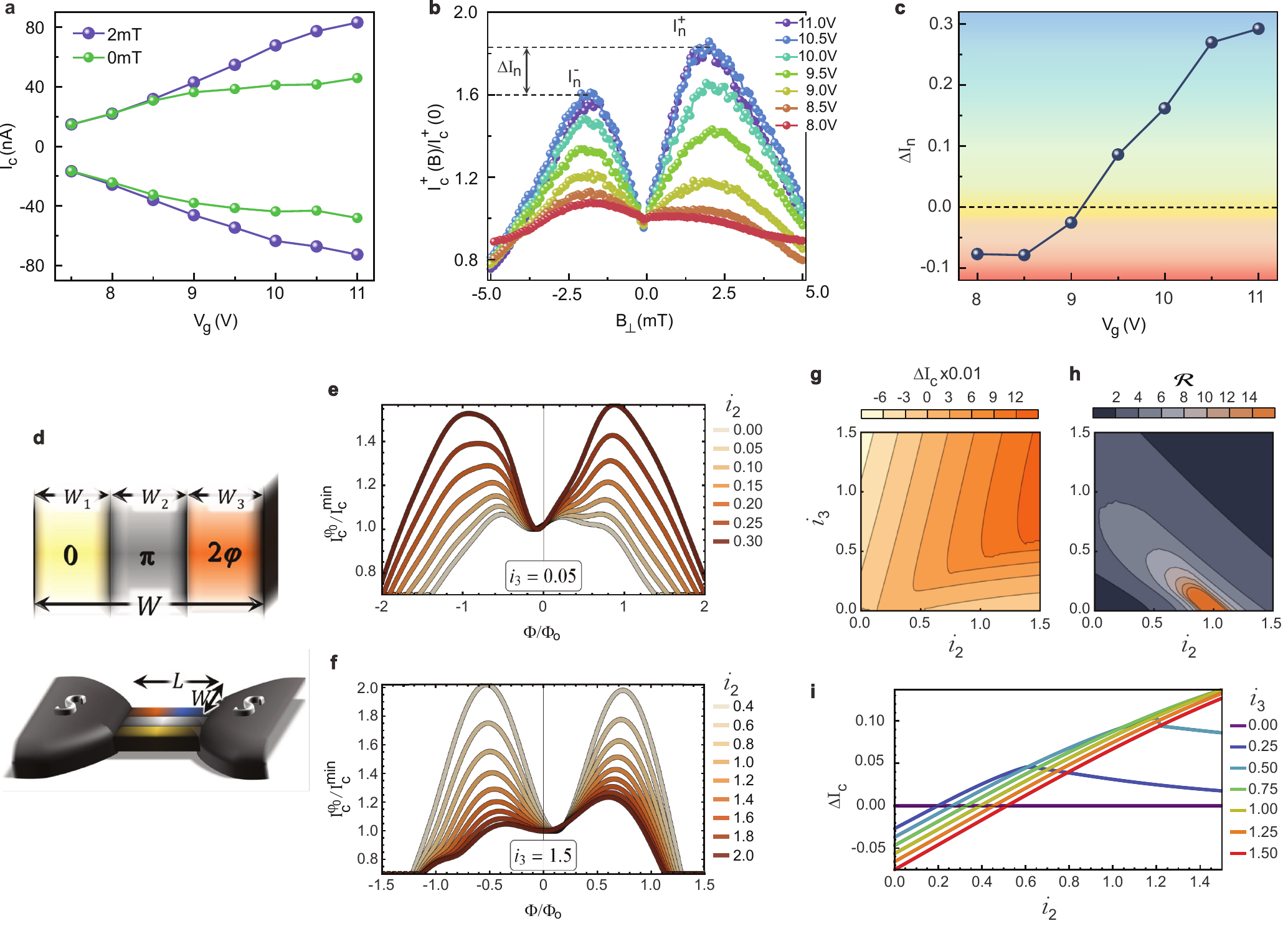}%
\end{center}
\vskip -0.5cm
\caption{\label{fig4} \textbf{Sign reversal, three channels model with intrinsic phases.} \textbf{a,} The evolution of the critical current estimated at constant magnetic fields $0\;\text{mT}$ and $+2\;\text{mT}$, with gate voltages between $11\;\text{V}-7.5\;\text{V}$. \textbf{b,} The normalized positive critical current $I_c^+(B)/I_c^+(0)$ \emph{vs} $B_\perp$ estimated from I-V \emph{vs} $B$ curves at different gate voltages. The dashed lines represent the value of current ($I_n$) at $\pm2\;\text{mT}$. \textbf{c,} The variation of $\Delta I_n=I_n^+(2\;\text{mT})-I_n^-(-2\;\text{mT})$ as a function of $V_g$. A clear crossing of the zero line takes place and this indicates the sign reversal at $\sim9.5\;\text{V}$. \textbf{d,} Illustration of the three-channel theoretical model, where the total width is $W = W_1+W_2+W_3$, with corresponding intrinsic phase drop of $\varphi_{in}=0$, $\pi$, and $\varphi_0$. \textbf{e,f} Magnetic flux dependence of the normalized critical current $I_c^{\varphi_0}(\Phi)/I_c^{\text{min}}$ (where $I_c^{\text{min}}=\text{\text{min}}\left [ I_c^{\varphi_0}(\Phi \sim 0) \right ]$) at fixed $i_3 = 0.05$ and $1.5$ and different values of $i_2$, and equal channel widths, $W_1=W_2=W_3=W/3$. Here, $i_2= I_{20}/I_{10}$ and $i_3= I_{30}/I_{10}$ indicate the ratios between the weights of the channels $\pi$ and 0 and between the channels $\varphi_0$ and 0, respectively. \textbf{g,} The variation of $\Delta I_c$ as a function of $i_2$ and $i_3$. Here, $\Delta I_c=\mathrm {max} \left [ I_c^{\varphi_0}(\Phi > 0)\right ] - \mathrm {max} \left [ I_c^{\varphi_0}(\Phi < 0) \right ]$. \textbf{h,} The ratio $\mathcal{R}$ of maximum/minimum of $I_c^{\varphi_0}(\Phi)$ as a function of $(i_2,i_3)$. \textbf{i,} The variation of $\Delta I_c$ as a function of $i_2$ at different values of $i_3$. }
\end{figure*}

The normal resistance, as shown in the Supplementary Fig. S13, does not depend on the magnetic field and thus cannot be the reason for the observed enhancement of $I_c$, as reported earlier in small magnetic fields~\cite{Chen-2009, Chen-2011}. Non-equilibrium quasiparticle poisoning has been recently discussed in relation to the enhancement of $I_c$, which is a result of improved thermalization due to the softening of superconducting gap with the effect of the magnetic field~\cite{Murani-2019}. The thermalization effect in nanowires is expected to be much stronger in narrower structures due to limited electron-phonon cooling power~\cite{Courtois-2008}. To address this issue, we measured two different kinds of uniform devices: single-connected (Supplementary Fig. S10a) and double-connected (rings) (Supplementary Fig. S10b), with different widths in the range of $w = 0.1\;\mu\text{m}-1\;\mu\text{m}$ for planar devices and $w = 0.1\;\mu\text{m}-0.3\;\mu\text{m}$ for ring structures. We see a systematic reduction of the amplitude of $I_c$ enhancement with reducing the width of the devices both for planar and ring-shaped devices. This confirms that the quasiparticle heating cannot be accounted for the $I_c$ enhancement. Moreover, the re-trapping current shows no increase with $B_\perp$, again corroborating against thermalization effects~\cite{Murani-2019} (Supplementary Fig. S11).

A completely different and more fundamental view of the enhancement of critical current by small perpendicular magnetic field is based on the interference of supercurrents flowing in channels with intrinsic phase shifts~\cite{Kemmler-2010}. In general, non-trivial phase relation of the superconducting order parameter appears when time or symmetry inversion is broken~\cite{Szombati-2016, Strambini-2020, Assouline-2019}. On the basis of symmetry arguments and considering the electronic structure of the LAO/STO interface, two physical scenarios emerge as the most relevant and plausible. The LAO/STO is a quasi-2D non-centrosymmetric superconductor and, due to confinement and electrostatic gating, the electronic structure is governed by strong inversion asymmetry. On the other hand, there have been experimental reports indicating the presence of ferromagnetic ordering at the LAO/STO interface co-existing with superconductivity providing an explicit source of time reversal symmetry breaking~\cite{Bert-2011}. Both these symmetries breaking may support anomalous Josephson channels in the superconducting phase. Furthermore, a key aspect of LAO/STO interface is given by the presence of multi-orbital degrees of freedom that hints for having more than one channel with different phase shifts, which in turn can be tuned by applied gate voltage. Since we do not have any experimental evidences of the presence of significant hysteresis in $I_c-B$ and $R_n-B$ (see supplementary Fig. S13) in our samples and symmetry by reversing the sign of both $I$ and $B_\perp$, it is plausible to focus on a model in which the inversion symmetry breaking is the primary source of an unconventional superconducting state associated with anomalous magnetic field dependence of the critical current. Here, the main challenge is to understand whether the experimental features can be accounted for within this scenario in a way that it is not much sensitive to the fine tuning of the model parameters.

In order to explain both the presence of a minimum of critical current at zero magnetic field, and a sign tunable asymmetry of $I_c(H)$, we propose a model assuming that the resulting zero magnetic-field critical current across the weak link is a superposition of three current channels having intrinsic phase drops with $\varphi_{in}=0$, $\pi$ and a $\varphi_{0}$ contributions, see Fig. 4d. Thus, the constituting currents can be expressed as following: $I_1=I_{10}\sin(\varphi)$, $I_2=I_{20}\sin(\varphi+\pi)$, and $I_3=I_{30}[\sin(\varphi)+\gamma \sin(2\varphi)]$. It is crucial that $\gamma$ is negative and has an amplitude relation that allows having a minimum of the Josephson energy at a non-zero phase bias $\varphi_{0}$ (details are reported in the supplemental information). This phase state is time-reversal invariant (i.e., there is no spontaneous current flow at $\varphi=0$), and the contributions can be deduced by assuming a multiband non-centrosymmetric superconductor with combination of singlet and triplet pairs, thus stemming only on the role played by inversion symmetry breaking and multi-orbital degrees of freedom.
The application of an external magnetic field modulates the amplitude of supercurrent as expected in a conventional Josephson junction. Assuming a junction of length $L$ and width $W$, the magnetic field dependence of the current for a given channel can be expressed as $I(H,\varphi)=J L\int_{0}^{W} \sin(2\pi \frac{HLx}{\Phi_{0}}+\varphi+\varphi_{in})$, with $J$ being the current density and $\Phi_{0}$ the unit of flux quantum. The critical current can be determined by summing up the contributions from all channels and maximizing the result with respect to the phase $\varphi$. The results of calculations are presented in Fig. 4e-i. The model is successfully capturing the following experimental key features qualitatively: i) the maximum of the $I_c$ that occurs at a non-zero value of the $B_\perp$ as shown in Fig. 4e,f ii) the minimum of $I_c$ at about $B_\perp = 0$, iii) the absence of parity in the pattern $I_c$ \emph{vs} $B_\perp$, i.e., by reversing the direction of the magnetic field ($B \rightarrow -B$) the critical current is not symmetric (Fig. 4e), and finally iv) the sign of the $I_c - B_\perp$ asymmetry, expressed through the difference between the $I_c$ at positive and negative applied field $B_\perp$, that can be reversed by applying an external gate voltage (Fig. 4f,g). Remarkably, the presence of the $\varphi_{0}$ channel is necessary to obtain the asymmetry and overall current-field profile, even without the fine tuning of the size of the regions corresponding to the $0-$, $\pi-$, and $\varphi_{0}$ channels. This implies that the result does not depend on the junction's geometry and the relative width of the supercurrent channels.

Concerning the origin of the Josephson components, the gate tunability of the supercurrent patterns clearly points to a substantial role of the orbital degrees of freedom in setting out a non-trivial pairing structure.
The gate voltage range, in our case, always corresponds to multi-band (d$_{xy}$ and d$_{xz,yz}$) occupation, as identified in the Hall effect (supplementary Fig. S14). Then, the $\pi$-channel can be linked to both inter-orbital spin-singlet states with superconducting order parameters with opposite sign~\cite{Scheurer-2015-2} or inter-orbital spin-triplet pairs~\cite{Fukaya-2018} and can be generally stable in the presence of inhomogeneous spatial conditions as well as inversion symmetry breaking. This implies that a time reversal invariant superconductor with local inter-orbital singlet and triplet pairs is compatible with the occurrence of independent $\pi$- and $\varphi_0$-channels which are gate tunable.

We note that our data are qualitatively similar to recently reported anomalous Josephson effect in proximity-induced $s_{+-}$ superconducting state in SnTe nanowires with ferroelectric domain walls~\cite{Trimble-2019}. In this work, however, the $\pi$-pairing leads to a time reversal broken state that in turn would support a spontaneous zero-field critical current with asymmetric magnetic field dependence and pronounced minimum at zero field but different patterns upon sign reversal of magnetic field and critical current. In our case, the application of a three-channels model with time-reversal symmetry breaking shows a lower degree of matching of the main features of the experimental results for inversion of the magnetic field orientation (details in the supplemental information) and lacks the symmetry for inversion of the critical current. On the basis of the above arguments, we deduce that a superconducting state which breaks time reversal symmetry is unlikely to occur in LAO/STO interface.

We conclude that inversion symmetry breaking is sufficient to account for an unconventional superconducting state with multi-channels that, in turn, are able to capture the hallmarks of the anomalous magnetic field behavior of critical current. In particular, the gate tunability of the observed effects highlights the role of inter-orbital pairing for the LAO/STO system which is dominantly active in the regime of the occupation of the d$_{xy}$ and d$_{xz,yz}$ bands and would account for the possibility to control the relative strength of the unconventional pairing channels.

Our results provide clear-cut evidences for the coexistence of multi-channel unconventional pairing in non-centrosymmetric oxide 2D superconductor and thus might open new routes for tailoring topological superconducting phases through precise control of band occupation and phase offsets in supercurrent channels.

\vskip 1cm
\textbf{\Large Methods}
\vskip 0.1cm
\textbf{Sample preparation} The main data presented in the article for 5uc-LaAlO$_3$ film grown on TiO$_2$-terminated STO substrates by ablating a single crystal LAO target using pulsed laser deposition ($\lambda = 248\;\text{nm}$) with fluence energy 1.0 J/cm$^{2}$ and Laser repetition rate $1\;\text{Hz}$. The substrate was heated to temperature $700^\circ\;\text{C}$ and film was deposited in an oxygen partial pressure p$_{O2}= 0.1\;\text{mbar}$. The epitaxial growth was monitored using in situ reflection high-energy electron diffraction (RHEED). The RHEED showed clear intensity oscillations confirming layer-by-layer growth. After the growth, film was annealed at 500 $^{0}$C in an oxygen pressure $\sim300\;\text{mbar}$ to avoid any oxygen deficiency which could lead to extrinsic conduction in the film.
\vskip 0.5cm
\textbf{Device patterning} The uniform bridge devices were patterned by following steps: (1) Ti/Au contacts ($5\;\text{nm}$ Ti and $100\;\text{nm}$ Au) were fabricated by lift-off technique and dc magnetron sputtering. This method provides low-ohmic contact resistance to the LAO/STO interface. (2) A thin layer ($\sim60\;\text{nm}$) of negative resist (ma-N2401) deposited on the LAO surface. (3) The resist mask was patterned by e-beam lithography (JEOL JBX-9300FS). (4) The sample was then plasma irradiated by a low energy Ar+ ion beam in an Oxford IonFab 300 Plus system using an inductively coupled plasma Ar+ source and 3-cm beam aperture. The sample was irradiated for $\leq$2 minutes by low energy Ar+ ion beam with beam energy of 150 eV and a current density of 0.03 mA/cm$^2$, that results in insulating interface in areas not covered by the e-beam resist. The irradiation process does neither result in the physical removal of the LAO film nor produces oxygen vacancies in the STO layer [1-3]. We monitored the resistance of the interface in real-time irradiation, and the process stopped when resistance increases up to $\sim$20 M$\Omega$ to prevent overexposure of the interface. The quality of nanopatterning was checked by measuring temperature-dependent sheet resistance before and after patterning, which concludes no change in the behavior in the temperature dependence of resistance R(T). In one device presented in Fig. 1a with bridge width 1$\mu$m, we fabricated a weak link of width d $\sim35\;\text{nm}$ by creating a small opening in positive resist (950k PMMA A2) using EBL. In this case, the sample was irradiated by low energy Ar+ ion beam for a relatively shorter time ($\leq$ 1 minute) than the first lithography to make sure that the resistance of the weak link remains in the quasi-metallic regime.
\vskip 0.5cm
\textbf{Transport measurements} The I-V characteristics and other transport measurements are performed in a dilution refrigerator (Oxford Triton) which equipped with combination of mu-metal shields and a superconducting lead shield protecting from the background DC magnetic field component, with a residual field of less than 100 nT. The external noise is suppressed by using twisted pairs of superconducting NbTi/Cu lines and combination with two stages of cryogenic filters. A Cu powder filters was installed at the base temperature stage of the refrigerator, together with low-pass RC filters with a cutoff frequency of 0.2 MHz at the 4 K stage. The noise is further filtered by using conventional EMI filters at room temperature. The out of plane magnetic field is applied by using a Helmholtz NbTi/Cu superconducting coil, whereas, in-plane magnetic field is applied using cylindrical NbTi/Cu superconducting coil capable of a maximum magnetic field of $25\;\text{mT}$. Room temperature low noise filter amplifiers were used for conventional pseudo-four-probe measurements of the I-V characteristics.
\vskip 0.5cm
\textbf{\Large Acknowledgements}\\
R.C. and C.G. would like to acknowledge useful discussion with F. Romeo. This research was funded by ERA-NET QUANTERA European Union’s Horizon H2020 project "QUANTOX" under Grant Agreement No. 731473 and Swedish Research Council (VR) grant number 2016-05256. E.L. acknowledges funding from the European Union’s Horizon 2020 research and innovation programme under the Marie Skłodowska-Curie grant agreement No 707404. M.C. acknowledges support by the project “Two-dimensional Oxides Platform for SPINorbitronics nanotechnology (TOPSPIN)” funded by the MIUR-PRIN Bando 2017 - grant 20177SL7HC. We also acknowledge support from the Swedish infrastructure for micro- and nanofabrication - MyFab.

\end{document}